\documentclass[pra,twocolumn,graphicx,amssymb,floatfix]{revtex4}
\usepackage{graphicx}
\begin{document}

\title{Particle path(s) through a nested Mach-Zehnder interferometer: Reply to Griffiths }

\author{L. Vaidman}
\affiliation{ Raymond and Beverly Sackler School of Physics and Astronomy\\
 Tel-Aviv University, Tel-Aviv 69978, Israel}

\begin{abstract}
   Griffiths [Phys. Rev. A 94, 032115 (2016)] analyzed, in the framework of consistent histories interpretation, the controversy regarding the approach to the past of a quantum particle introduced by Vaidman [Phys. Rev. A 87, 052104  (2013)]. I argue that Griffith's criticism of my approach using analysis of experiments with weak probes is unfounded.
       \end{abstract}
\maketitle

In the abstract Griffiths \cite{Grif}  writes:
\begin{quote}
  Possible paths of a photon passing through a nested Mach-Zehnder interferometer on its way to a detector
are analyzed using the consistent histories formulation of quantum mechanics, and confirmed using a set of
weak measurements (but not weak values). The results disagree with an analysis by Vaidman [Phys. Rev. A 87,
052104 (2013)], and agree with a conclusion reached by Li et al. [Phys. Rev. A 88, 046102 (2013)].
\end{quote}
I argue that the Griffiths  analysis of  weak measurements is conceptually incorrect.

 Griffiths  considered  nested Mach-Zehnder interferometer
tuned such that there is a destructive interference of the inner interferometer towards $E$,  see Fig.~1a. A particle is emitted at $S$ and is detected at $D_1$. The question is: ``Where was this particle at intermediate times?'' In Section II B Griffiths writes that the ``correct'' answer is ``path $A$'', see Fig.~1b. This is the claim of Li {\it et al.}, \cite{LiCom}, while I argued that the particle was in path $A$ {\it and} inside the inner interferometer in paths $B$ and $C$ \cite{past,pastReply}, see Fig.~1c.  In Section III D, however,  Griffiths writes that for a particular choice of the reflection coefficients  (his Eq. (22)) and for a particular consistent family, ``path $C$'' is a correct answer too, so that there is no unambiguous answer to the question.

The consistent histories approach does not allow to say that it is both in $A$ and in $C$ as suggested by my approach, but this is not really a contradiction. Griffiths agrees with me that standard quantum mechanics does not provide an answer to the question: ``Where was the particle at the intermediate time?'' If we want to answer this question, we have to add some principle(s). In his paper Griffiths explains  why in this particular situation consistent histories approach does not provide an unambiguous answer. The consistency rules are build to avoid paradoxes in classical way of thinking about quantum experiments.

Our disagreement is about the
  analysis of  weak measurements in this setup. The analysis appears in Section~V where Griffiths evaluates the effect of the pre- and post-selected particle on weakly coupled  probes    placed  in various paths of the interferometer, see Fig.~1d. He correctly points out that it does not matter if the  probes are qubits or Gaussian pointers used in my analysis \cite{JPA}.
   He summarises his results in his Eq. (35) and admits that that to the first order, the probes show the picture I advocate: the particle leaves trace in $A$, $B$ and $C$. But then he writes: ``However, the coincidences, two or more probes triggered during a single run, agree with the alternative explanation given in \cite{LiCom}.''

  First, let me repeat  my claim. I {\it define} that the pre- and post-selected particle was where it left a weak trace. The trace is a local change of environment, which, in particular, might be the change of quantum states of local probes in the paths of the interferometer. Given the setup, it can be calculated using standard rules of quantum mechanics. In our case, if we put similar probes in every path of the interferometer, we obtain that  the change of their quantum states, estimated, say, as the Bures angles relative to their states without the particle, present, to the first  order in the strength of the coupling,  in probes $S,A,B,C$, and $F$. It is uncontroversial that the the particle was in $S$ and in $F$. Therefore, the trace criteria tells us that it was also in $A,B$, and $C$. The principle, the particle was where was the trace, together with quantum calculation of the trace, tells us that the particle was on path $A$, but also inside the inner interferometer in paths $B$ and $C$, without being in paths $D$ and $E$, see Fig.~1c.

\begin{figure}[h]
\begin{center}
 \includegraphics[width=5.6cm]{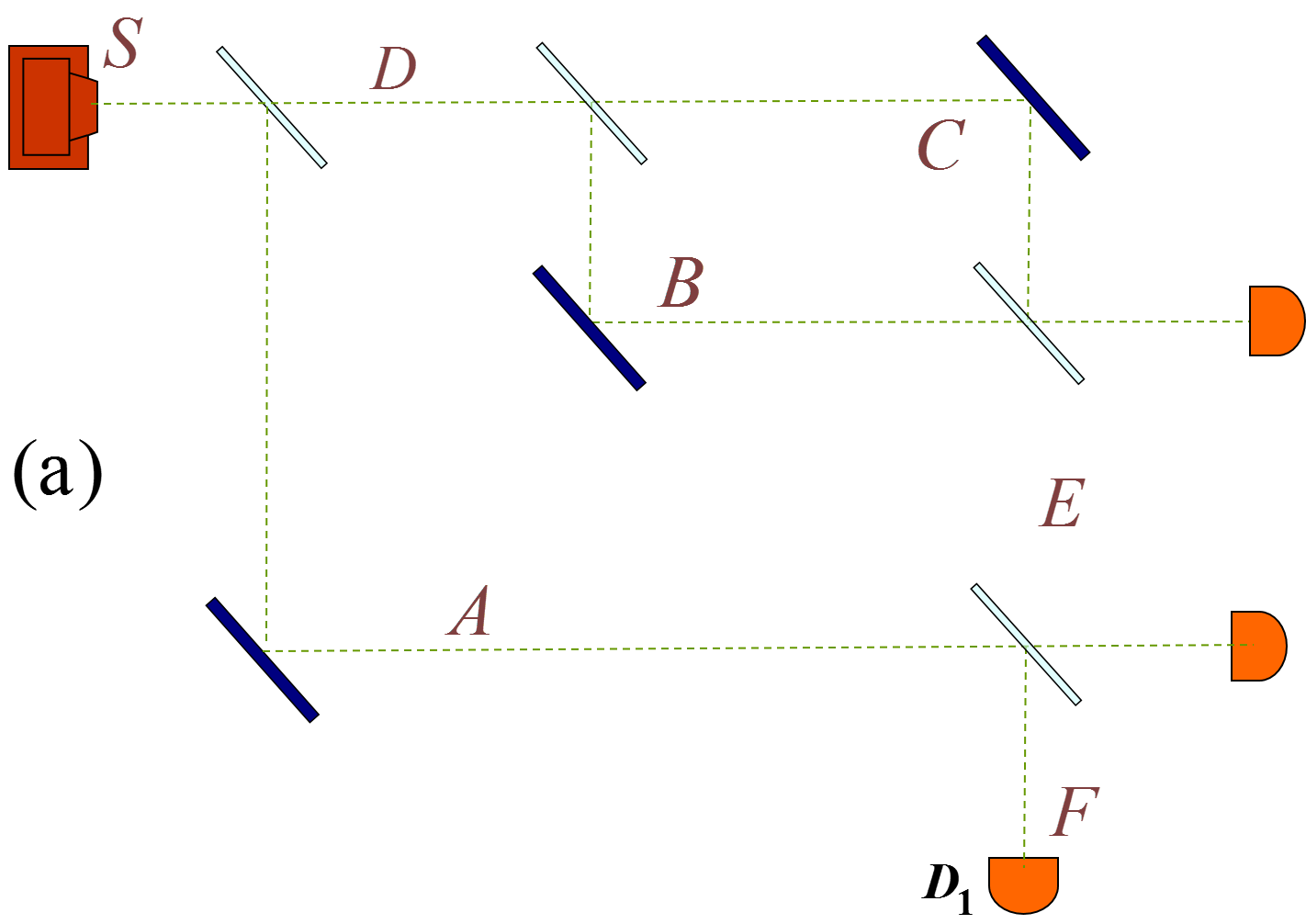}  \includegraphics[width=5.6cm]{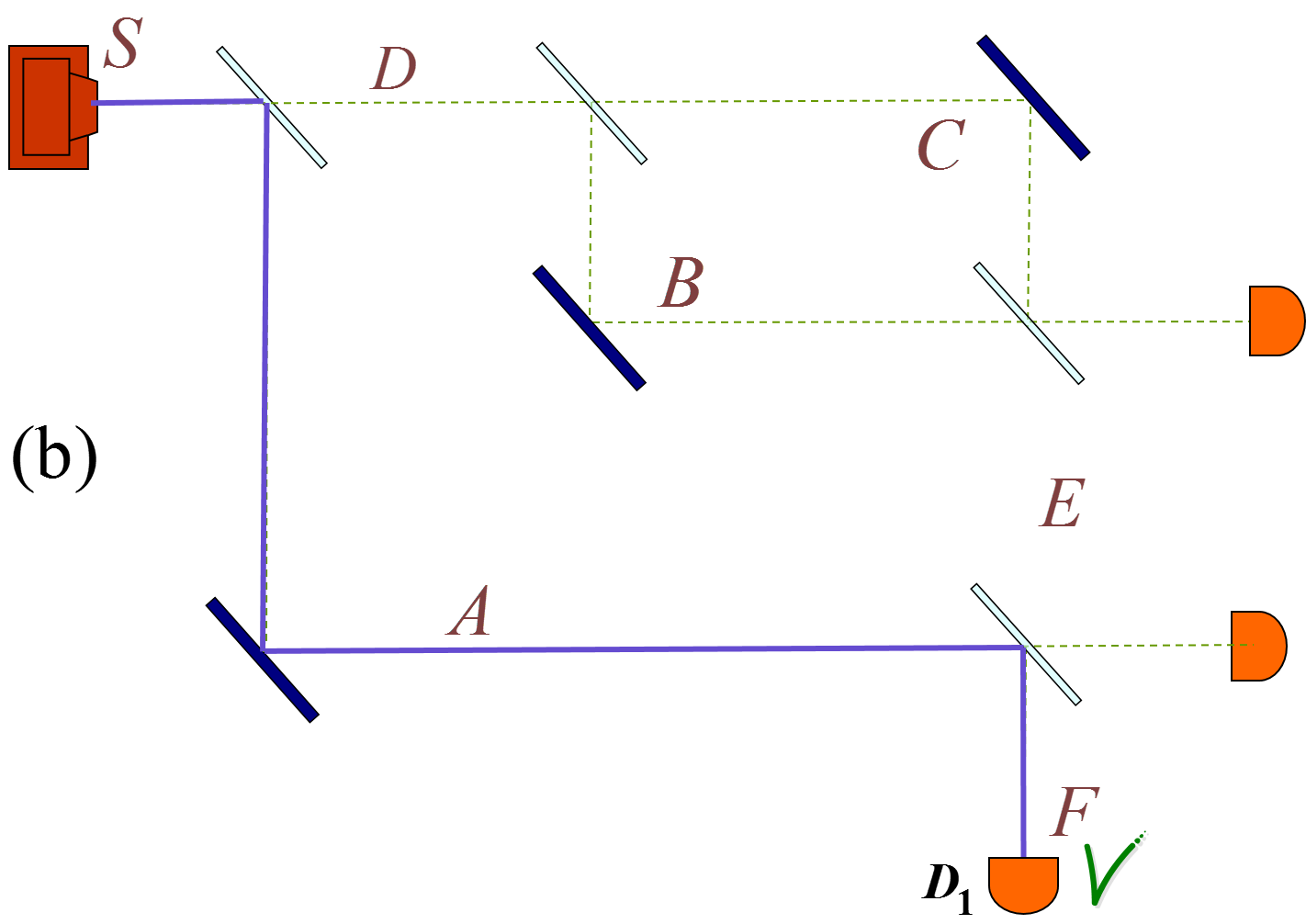} \includegraphics[width=5.6cm]{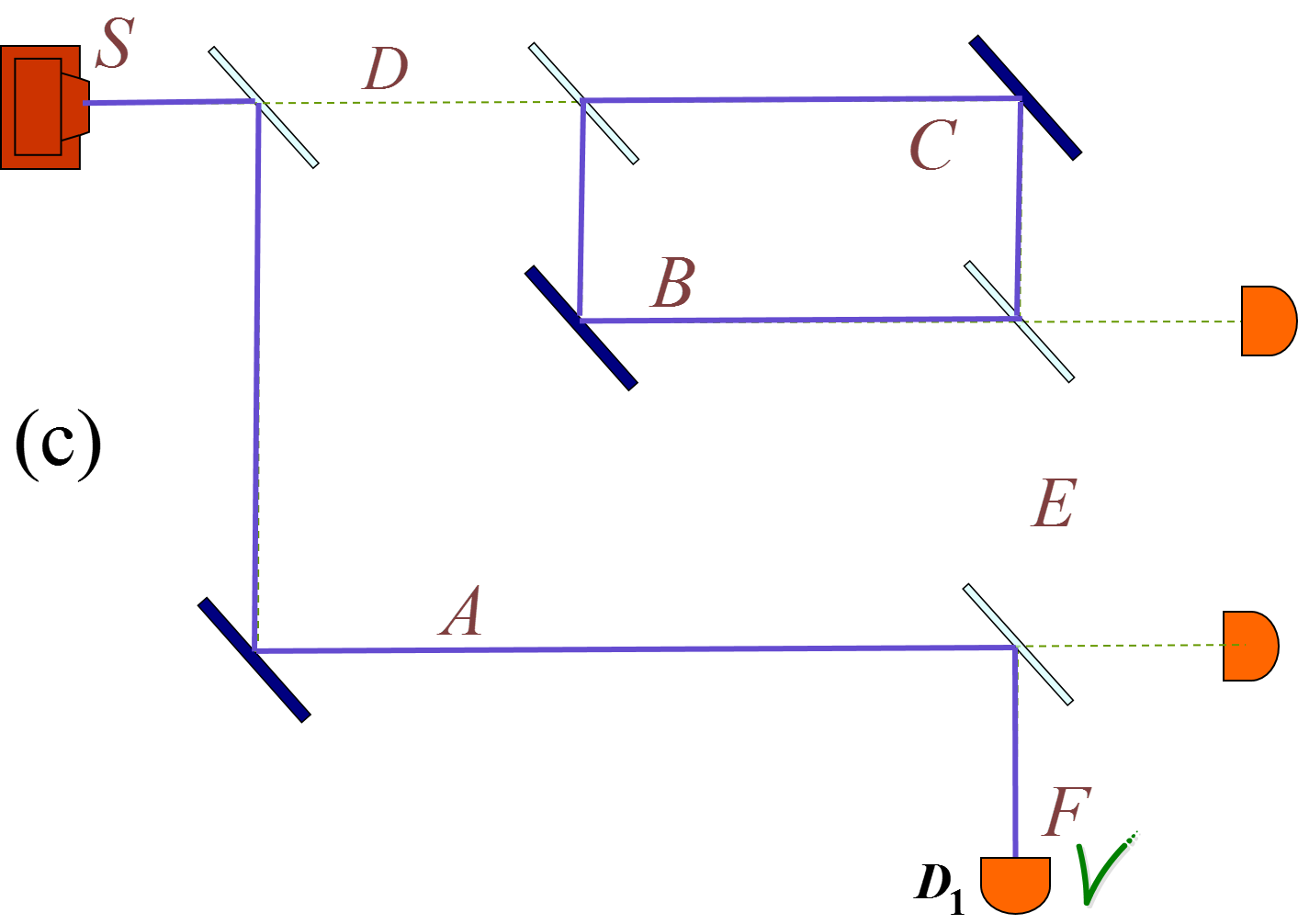} \includegraphics[width=5.6cm]{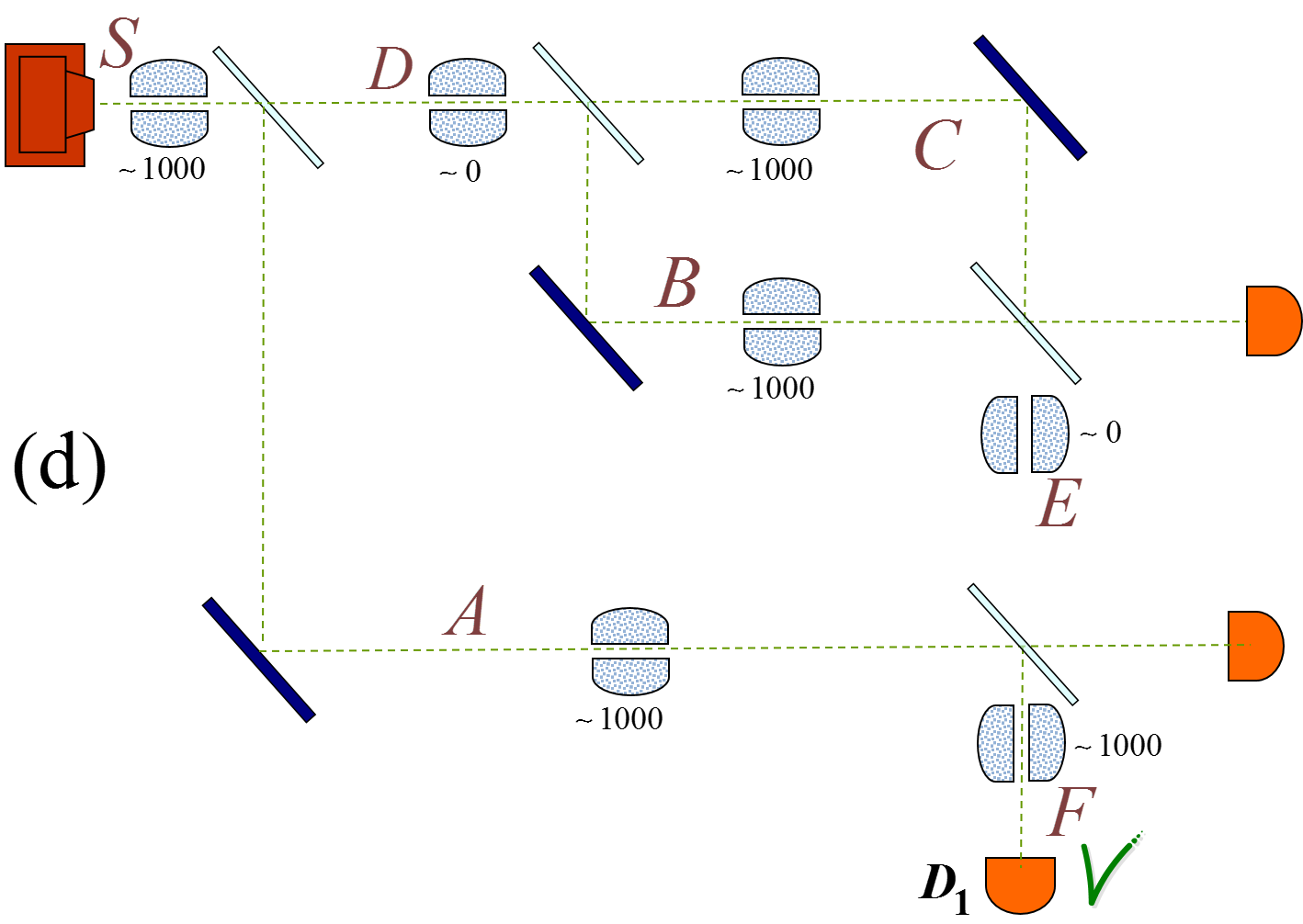}\end{center}
\caption{ a). Nested Mach-Zehnder interferometer.  The inner interferometer is tuned such that there is a destructive interference  towards $E$.  b) The location of the particle detected by $D_1$ inside the interferometer according to Li {\it et al}. \cite{LiCom}. c)  The location of the particle detected by $D_1$ inside the interferometer according to Vaidman \cite{past}. d) Counts of the triggered weak probes in paths of the interferometer experiment performed on the ensemble of $10^7$ particles started at source $S$ and post-selected in detector $D_1$.}
\end{figure}

 Griffiths, instead of analyzing the states of the probes,  relies on the clicks of additional measurement devices which observe the quantum state of the probes. He writes:
\begin{quote}
After a given run is finished each probe can itself can be subjected to a strong measurement in the $|0\rangle$, $|1\rangle$ basis to determine its value. A probe state $|1\rangle$ indicates that the particle was in that channel (or in $B + C$ for probe $w$), but if the state is $|0\rangle$ one learns nothing: The particle might have been in the channel, but if so, it left no trace.
\end{quote}
I disagree that ``no click'' provides no information: likelihood of the presence of the particle near the silent probe is reduced. But the next sentence is even more problematic:
\begin{quote}
Note that the process
of measuring the probes, which takes place after the particle
has completed its path through the trajectory, has no effect
upon that trajectory, since the future does not influence the
past; instead, the measurement yields information about the
state of affairs at the earlier time.
\end{quote}

Griffiths does not claim that interaction with probes causes collapse. It is the ``strong measurement'' of the probe which causes collapse (or splitting the worlds for physicists accepting the many-worlds interpretation). Of course, the state of the particle is changed (collapsed) by the measurement.  If a strong measurement of the probe finds that the state of the probe has changed, it collapses the state of the particle to the path of the probe. This event represents a strong measurement which tells us that the particle in {\it this} particular run took a particular path. It strongly disturbs the particle (as Griffiths admits) and therefore, the definite statement about where was {\it this} particle does not help to know where were undisturbed  pre- and post-selected particles inside  the interferometer.

If we run the experiment once and it happens that we got a click in one of the probes, or, even three clicks, in $D$, $B$, and $E$ (which has very low probability), ``the state of affairs at earlier time'' has changed: we know that this particle had trajectory $SDBEF$. Before the clicks we had no reason to claim this,  and in the framework of standard quantum mechanics without hidden variables, given only pre-selection in  $S$ and post selection in $F$, there is nothing which can carry this information.

The runs with clicks tell us where were the particles in these particular runs, but they also tell us that this is not reliable information about the location of particle in other runs. Even one click destroys the interference. When we  have ``the coincidences, two or more probes triggered during a single run,'' we do not get reliable information about the position of the particle in  runs without clicks. The runs with coincidences, which according to Griffiths contradict my approach, are not relevant.

One can see that coincidences with triggered probes $B$ (or $C$) are not relevant for the interference experiment which is discussed also from the fact that the probability  of the coincidence events do  not depend on the phase tuning of the inner interferometer. The special properties of the experiment crucially depend on the tuning to destructive interference towards $E$. Statistics of single events does depend on the tuning of the interferometer.

The runs with clicks are important, but not as experiments which occasionally show where was every particle in the ensemble. The runs with clicks are part of  the tomography experiment of the states of the probes in the pre- and post-selected ensemble which includes, importantly, the runs without clicks. The very rare runs with coincidences play a very minor role (if any) in this tomography experiment.

Griffiths considers the probability of triggering the probe, given that the particle is present, of the order of $10^{-4}$. Let us also consider the interferometer with equal intensity of beams in paths $A$, $B$ and $C$ corresponding to the Griffiths conditions expressed in his Eq. (22). In a realistic run with, say, $10^{7}$ particles, we get around thousand clicks in  the probes $S$, $A$, $B$, $C$, and  $F$, see Fig.~1d. Most probably, no clicks in probes $D$ and $E$. (The probes $D$ and $E$ have approximately  probability $0.2$ for  a click and it must be in coincidence with a click in $B$ or $C$.)

Like in any quantum experiment, the click in a particular run does not tell us what was the quantum state of the system.  The coincidence events do not tell us where were the particles in the vast majority of cases without clicks.  Only the full statistics gives the relevant information. The statistics of clicks of weak probes of all runs of the experiment supports my picture.

Another criticism of Griffiths is that an experiment with the probe $w$ which indicates the presence of the particle inside the inner interferometer without  distinguishing between the arms $B$ and $C$ will have no clicks, apparently  in contradiction with my claim that the particle was inside the inner interferometer. Yes, there will be no clicks in this probe. The reason for this, however, is not that the particle was not inside the interferometer, but that the probe was not able to detect it.

Consider an analogy. A  device, sensitive to an electric field, is placed in the center of the interferometer with charged particles. It clicks when a particle is present in one of the arms due to the electric field of the particle. However, when two particles are present, one in each arm, the electric field in the center vanishes and the device does not click.

In the current experiment, the setup is such that the effect of the particle in each arm of the inner interferometer, if it present only there, is identical. The opposite signs of the effects on the probe of the particle being in $B$ and the particle being in $C$ are due to a subtle interference phenomena.
 Indeed,  the setup  of the experiment with Griffiths condition (his Eq. (22)) corresponds to the pre-selection of the state $\frac{1}{\sqrt 3} (|A\rangle+ |B\rangle+|C\rangle)$ and post-selection of the state $\frac{1}{\sqrt 3} (|A\rangle - |B\rangle+|C\rangle)$. Consider the experiment with the probe $w$ in which the interaction Hamiltonian is changed such that the coupling with the particle is present only when it is in path $C$. After the post-selection, i.e., the detection of the particle in $D_1$, the state of the probe $w$ will be (cf.  Eq. (27) of Griffiths)
  \begin{equation}\label{C}
    \sqrt{1-\epsilon}|0\rangle_w +  \sqrt{\epsilon}|1\rangle_w.
  \end{equation}
If, instead, the interaction is switched on in $B$  and switched off in $C$, the final state of the probe $w$ will be
   \begin{equation}\label{B}
    \sqrt{1-\epsilon}|0\rangle_w - \sqrt{\epsilon}|1\rangle_w.
  \end{equation}
 When the interaction is switch on both in $C$ and in $B$, after the post-selection, the state of the probe is $|0\rangle_w$. So, the probe $w$, which interacts in a nonlocal way both with arm $C$ and arm $B$, will not show the presence of the particle in the inner interferometer. However, the relevant probes, which are real physical effects of local environment in arms $B$ and $C$, will record  the presence of the particle.

  Griffiths (and others)  might argue about strengths and weaknesses of my principle: the pre- and post-selected particle was where it left a weak trace. However, his calculations of weak measurements  do not refute my claim about the form of the trace in this nested interferometer.

This work has been supported in part by the Israel Science Foundation Grant No. 1311/14,
the German-Israeli Foundation for Scientific Research and Development Grant No. I-1275-303.14.

\end{document}